\documentclass[twocolumn,showpacs,preprintnumbers,amsmath,amssymb,pre]{revtex4-1}
\usepackage{graphicx}% Include figure files
\usepackage{dcolumn}% Align table columns on decimal point
\usepackage{bm}% bold math
\usepackage{xcolor}

\begin{document}

\title{Inhibited pattern formation by asymmetrical high voltage excitation in nematic fluids}

%\authors
\author{P\'{e}ter Salamon, N\'{a}ndor \'{E}ber, Bal\'{a}zs Fekete, and \'{A}gnes Buka}

\affiliation{Institute for Solid State Physics and
Optics,\\ Wigner Research Centre for Physics, Hungarian Academy of Sciences,\\ H-1525 Budapest, P.O.B.49,
Hungary}

\date{\today}% It is always \today, today,
             %  but any date may be explicitly specified

\begin{abstract}

In contrast to the predictions of the standard theory of electroconvection (EC), our experiments showed that the action of superposed ac and dc voltages rather inhibits pattern formation than favours the emergence of instabilities; the patternless region may extend to much higher voltages than the individual ac or dc thresholds. The pattern formation induced by such asymmetrical voltage was explored in a nematic liquid crystal in a
wide frequency range. The findings could be qualitatively explained for the conductive
EC, but represent a challenging problem for the dielectric EC.
\end{abstract}

\pacs{61.30.Gd, 47.54.-r, 89.75.Kd }

\maketitle
\section*{Introduction}

Instabilities in nonlinear dynamical systems can lead to formation
of patterns \cite{Cross1993}. In fluids, patterns are often
associated with vortex flow induced by various driving forces such
as temperature gradient (Rayleigh-B\'enard convection
\cite{Ahlers2009,Ahlers2012,Petschel2013,duPuits2014}), shear
(Taylor-Couette flow
\cite{Huisman2012,Duguet2013,Huisman2013,Deguchi2014},
Kelvin-Helmholtz instability \cite{Kuramitsu2012,Hurricane2012}),
or electric field (electroconvection (EC) \cite{Bukabook1996}).
Anisotropic fluids are especially convenient to study some general
features of dynamical systems, as they prone to show easily
observable convective patterns in applied electric fields
because of their optical anisotropy.

Electroconvection in nematic liquid crystals \cite{Bukabook1996}
can be induced by both direct (dc) and alternating (ac) voltages
in the same compound. In this paper, we show that the application
of asymmetrical voltages corresponding to the superposition of a
dc and a sinusoidal ac signal can inhibit the formation of
patterns, even if the ac and dc components are an order of
magnitude higher than the threshold voltages of the purely ac or
purely dc induced electroconvection.

Nematic liquid crystals are mostly composed of elongated molecules
with their long molecular axes fluctuating around an average
direction, the director $\mathbf{n}(\mathbf{r})$
\cite{deGennesbook2001}. Due to their uniaxial symmetry, nematic
materials can be characterized by two independent dielectric
constants measured with electric fields parallel or perpendicular
to the director ($\varepsilon_{||}$ and $\varepsilon_{\perp}$,
respectively). A positive or negative dielectric anisotropy
$\varepsilon_{a}=\varepsilon_{||}-\varepsilon_{\perp}$ allows to
align the director parallel with or perpendicular to the electric
field, respectively \cite{deGennesbook2001}.

Liquid crystals are studied and used mostly in thin (5-20 $\mu$m)
films, sandwiched between glass plates with transparent electrodes
providing an electric field along the cell normal. A proper
treatment can ensure strongly anchored, homogeneous director
alignment at the surfaces. In a planar cell, the homogeneous
director lies in the cell plane. Due to the electric field applied
perpendicular to the initial director, if $\varepsilon_a>0$, an
instability occurs at a critical voltage $U_{cF}$ leading to a
homogeneous director deformation, called the Freedericksz
transition.

The electric Freedericksz transition can be induced by dc
($U_{dc}$) as well as by sinusoidal ac ($U_{ac}$) voltages
of frequency $f$ \cite{Stewartbook2004}. In the case of an
asymmetrical driving, the applied voltage is described by $U =
U_{dc} + \sqrt{2} U_{ac} sin(2 \pi f t)$. The onset of the
instability can be achieved by different combinations of the two
control parameters, $U_{dc}$ and $U_{ac}$, characterized by a
frequency independent threshold curve: a quarter circle in the
$U_{ac}$ - $U_{dc}$ plane given by $U_{cF}^2 = U_{dc}^2+U_{ac}^2$.
Inside this curve the system is in its homogeneous basic state;
outside the initial planar state is deformed.

If $\varepsilon_a<0$, the electric field exerts a stabilizing
torque on the director, however, an instability can still take
place leading to a periodic director deformation by convection,
governed mostly by the Carr-Helfrich mechanism: spatial director
fluctuations lead to space charge separation due to the
conductivity anisotropy $\sigma_{a}=\sigma_{||}-\sigma_{\perp}$
($\sigma_{||}$ and $\sigma_{\perp}$ are the conductivities
measured with an electric field parallel and perpendicular to the
director, respectively); the Coulomb force induces flow forming
vortices due to the constraining surfaces; the flow exerts a
destabilizing torque on the director \cite{Bukabook1996}. Above a critical voltage
$U_c$, the fluctuations do not decay, but grow to a macroscopic
pattern of convection rolls characterized by a critical wave
vector $\mathbf{q_c}$. The typical case of electroconvection
(standard EC) can be observed in a planar cell filled with a
nematic liquid crystal with $\varepsilon_{a} <0$ and $\sigma_{a}
>0$. The resulting patterns correspond to a spatially periodic
system of convection rolls that appear as dark and bright stripes
perpendicular (or oblique) to the initial director in a
microscope.

Different modes of EC can be realized at the onset of the
instability depending on the frequency. Typically at high $f$, the
\textit{dielectric mode} is present; then by decreasing $f$, a
transition to the \textit{conductive mode} occurs at the crossover
frequency $f_c$ \cite{Bukabook1996}. In cells of typical thickness, this transition
is easily observable due to the largely different $\mathbf{q_c}$
of the two modes.

A comprehensive theoretical description of the different pattern forming
modes in nematic liquid crystals is provided by the so-called standard
model of EC (SM) \cite{Bodenschatz1988}, which has recently been improved
by including flexo-electricity (extended SM)\cite{Krekhov2008}.
It combines the equations of nematohydrodynamics with those of electrodynamics,
while assuming ohmic electrical conductivity. The extended SM provides $U_c(f)$,
$\mathbf{q}_c(f)$ and the spatio-temporal dependence of the
director $\mathbf{n}(\mathbf{r},t)$ at onset, in agreement with
experiments. For EC, the equations exhibit solutions with 3
different time symmetries. One is found at dc driving, where
$\mathbf{n}(\mathbf{r})$, the flow field $\mathbf{v}(\mathbf{r})$,
and the charge field $\varrho_e(\mathbf{r})$ are static. The other
two occur at ac voltage excitation: in the conductive mode, the
director and the flow is stationary in leading order (if $f$ is
much larger than the inverse director relaxation time
$\tau_d^{-1}$), so the time average of the director tilt over the
driving period is nonzero ($\langle n_z(t) \rangle \neq 0$) while
space charges oscillate with $f$; in the dielectric mode,
$\varrho_e(\mathbf{r})$ is stationary while
$\mathbf{n}(\mathbf{r})$ and $\mathbf{v}(\mathbf{r})$ oscillate
with $f$ (thus $\langle n_z(t) \rangle = 0$).

Pattern formation may also occur at superposing two electric signals.
The behaviour of EC has been studied at adding sinusoidal or square wave
voltages of two distinct frequencies ($f_1 <f_2$, $f_2$ being a multiple
of $f_1$) and a nontrivial threshold variation and a reentrant pattern
forming behaviour was found \cite{Heuer2006,Pietschmann2010}.
Here we report on the mostly unexplored case of superposing ac and
dc voltages.
As standard EC may occur at pure dc as well as at pure ac
excitation, one might expect it to arise at a
combined, asymmetrical driving
too, just like in the case of the electric
Freedericksz transition. We raised the questions, how this nonlinear
dynamical system responds to the asymmetric excitation and which kind of
pattern morphologies will occur. The
consequences of the asymmetric driving in standard EC are
mostly unexplored experimentally and are
challenging also from the theoretical point of view; the solution may
not be obtained as a simple superposition of the three modes of
different time symmetries mentioned above. Recent studies have
shown that the extended SM can well describe the limits of
stability in the $U_{ac}-U_{dc}$ plane for the conductive EC if $f$
is sufficiently low ($f\ll f_c$); then the threshold curve is
similar to that of the Freedericksz transition \cite{Krekhov2014}.
Our present experimental work aimed to
provide
a more comprehensive
study on a broad range of $f$. It will be shown that the onset
behavior of the system is qualitatively different depending on
the frequency.

\section*{Experimental}

Our experiments were carried out using the nematic mixture Phase
5. It exhibits $\varepsilon_{a} <0$ and $\sigma_{a} >0$ and is a
widely used material for studying standard EC \cite{Eber2012}.
When applying ac voltage driving, depending on $f$, both
dielectric and conductive EC can be observed. We used $d=10.8$
$\mu$m thick cells (E.H.C. Co., Japan) at the temperature of
$T=30\pm0.05$ $^{\circ}$C for the measurements. Applying dc
voltage, our planar samples showed EC. The patterns were observed
in a polarizing microscope using the shadowgraph technique
\cite{Rasenat1989,Trainoff2002,Pesch2013};
the EC thresholds were determined in the
$U_{ac}-U_{dc}$ plane at different frequencies of the ac signal.
The searching for patterns were done at a fixed ac-voltage varying the dc-component.

\section*{Results and discussion}

\begin{figure}[!h]
\begin{center}
\includegraphics[width=8cm]{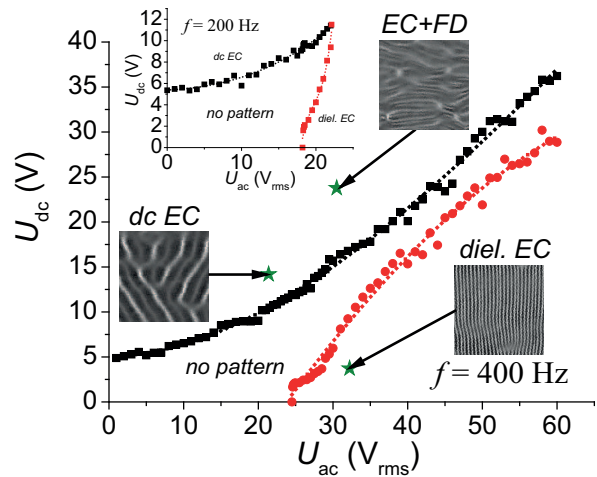}
\end{center}
\caption{(Color online) Morphological phase diagram of a Phase 5
sample at $f=400$ Hz, and at $f=200$ Hz (inset from
\cite{Krekhov2014}). The dashed lines, as a guide
for the eyes, indicate the trends of the stability limit curve.
Stars indicate those $U_{ac}$, $U_{dc}$ combinations where the
images covering an area of 52 $\mu$m $\times$ 52 $\mu$m were
taken. The initial director lies along the horizontal direction.}
\label{fig:1}
\end{figure}

In Fig.~\ref{fig:1}, the threshold curves showing the limits of
stability in the $U_{ac}-U_{dc}$ plane are presented for $f=400$
Hz. They exhibit a strikingly different behavior compared to the
quarter-circle of the Freedericksz transition. The threshold
curves starting from $U_{dc}=0$ V and $U_{ac}=0$ V form two
branches (the ac- and dc-branches, respectively), which do not
cross each other in the available voltage range. At $U_{dc}=0$ V,
$U_{ac}\gtrsim 24$ V, regular dielectric EC rolls are seen that
lie normal to the initial director. Following the threshold curve
on the ac-branch, the morphology remains the same, but
surprisingly, with increasing $U_{dc}$ along the curve, $U_{ac}$
also becomes higher; the curve bends away from the origin towards
higher $U_{ac}$, $U_{dc}$ values. If $U_{ac}=0$ V, dc EC can be
observed as a roll structure oblique to the initial director.
Following the threshold curve on this dc branch, with increasing
$U_{ac}$ EC appears at higher $U_{dc}$, and the slope remains
positive. We note that at $U_{dc} > 23$ V, besides EC, stripes
parallel to the director also appear in patches; they are
attributed to flexoelectric domains (FDs
\cite{Eber2012,Bobylev1977,Krekhov2011,Salamon2013,Eberbook2012}).

The two branches of the stability limit in Fig.~\ref{fig:1}
correspond to patterns with different morphologies and
significantly different wave numbers. In the narrow channel
between the two branches no patterns could be detected.
Surprisingly, the system remains there in the basic, undistorted
state despite of the high voltages applied. For example, at
$U_{dc}=32$ V and $U_{ac}=55$ V, the dc and ac voltage components
are more than 6 and 2 times larger than the corresponding
thresholds for purely dc and purely ac driving, respectively. In
the case of purely ac or dc driven EC, at voltages so much above
the thresholds, the convection would already be in the turbulent
regime. Our findings thus indicate that using a signal with
properly adjusted asymmetry can result not only in the suppression
of undesirable turbulence but also in complete inhibition of
pattern formation. An unusual sequence of morphologies can be
obtained by varying one voltage component while the other is
kept constant. For example, at fixed $U_{ac}=40$ V, with no dc component,
the convection is turbulent. Increasing $U_{dc}$, the system behaves less
and less overdriven; it shows regular patterns at $U_{dc}\approx 16$ V, then
if $16 \mathrm{\,V}\lesssim U_{dc}\lesssim 20 \mathrm{\,V}$, there is no
pattern at all. Applying higher $U_{dc}$, electroconvection sets in again,
and becomes turbulent at high values of $U_{dc}$.

The inhibition of pattern formation at combined ac and dc driving
holds also at higher frequencies. Decreasing $f$, however, leads
to a qualitatively different behavior. At $f=200$ Hz (inset in
Fig.~\ref{fig:1}), the pattern morphologies in the $U_{ac}-U_{dc}$
plane at onset are similar, but now the ac and dc branches cross
each other; the pattern-free channel closes at some voltages,
where a morphological transition occurs between the conductive and
dielectric roll structures \cite{Krekhov2014}.

\begin{figure}[!h]
\begin{center}
\includegraphics[width=8cm]{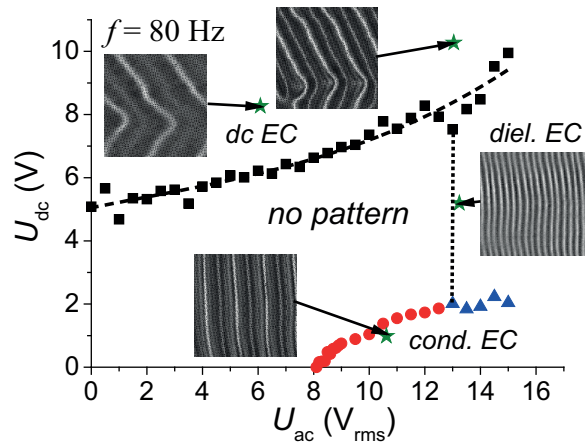}
\end{center}
\caption{(Color online) Morphological phase diagram of a Phase 5
sample at $f=80$ Hz. The dashed lines, as a guide
for the eyes, indicate the trends of the stability limit curve. The dotted line shows the boundary between the patternless basic state and the region of dielectric electroconvection.
Stars indicate those $U_{ac}$, $U_{dc}$ combinations where the
images covering an area of 52 $\mu$m $\times$ 52 $\mu$m were
taken. The initial director lies along the horizontal direction.}
\label{fig:2}
\end{figure}

The purely ac driving at $f = 80$ Hz yields conductive EC. The
ac-branch of the stability limit curve in Fig.~\ref{fig:2}
exhibits a positive slope in the $U_{ac}-U_{dc}$ plane until
$U_{dc}=2$ V, where the conductive roll structure crosses over to
the dielectric one, indicated by a large increase of the wave
number. At this morphological transition, the slope of the curve
also changes abruptly: for dc voltages $U_{dc}>2$ V the threshold
curve is characterized by an unaltered $U_{ac}$ component until
the crossing with the dc-branch (see dotted line in Fig.~\ref{fig:2}). There an additional morphological
transition occurs between the dielectric and the dc EC modes,
shown again by a significant change in the wave number.

\begin{figure}[!h]
\begin{center}
\includegraphics[width=8cm]{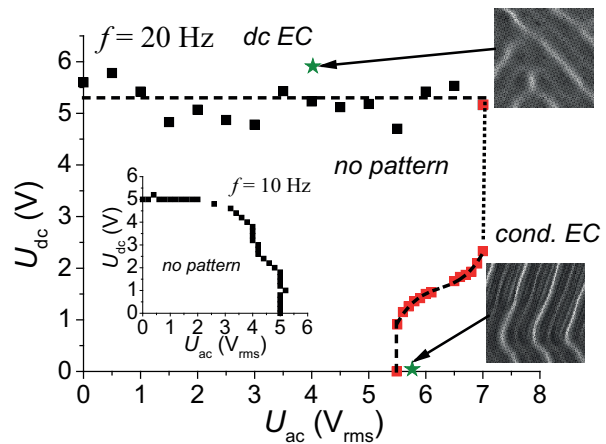}
\end{center}
\caption{(Color online) Morphological phase diagram of a Phase 5
sample at $f=20$ Hz, and at $f=10$ Hz (inset from
\cite{Krekhov2014}). The dashed lines, as a guide
for the eyes, indicate the trends of the stability limit curve. The dotted line shows the boundary between the patternless basic state and the region of conductive electroconvection.
Stars indicate those $U_{ac}$, $U_{dc}$ combinations where the
images covering an area of 52 $\mu$m $\times$ 52 $\mu$m were
taken. The initial director lies along the horizontal direction.}
\label{fig:3}
\end{figure}

At even lower frequencies, a dc-voltage-induced transition to the
dielectric EC does not occur; as a consequence there is no
dramatic change in the critical wave number along the stability
limit curve in the $U_{ac}$ - $U_{dc}$ plane. Nevertheless,
depending on the frequency, the system can show different
characteristics. At $f=20$ Hz (see Fig.~\ref{fig:3}), the
ac-branch shows mainly a positive slope that results in a larger
pattern-free area compared to the expected quarter-ellipse-shaped
threshold curve, found earlier at $f=10$ Hz \cite{Krekhov2014}
(see inset in Fig.~\ref{fig:3}).

Recently, the critical voltages and wave numbers were calculated
at the onset of EC induced by superposed ac and dc voltages
\cite{Krekhov2014}. Both analytical and numerical calculations
predicted that the dc-branch has positive slope in the
$U_{ac}$-$U_{dc}$ plane at higher frequencies and negative slope
at lower frequencies. This is in good agreement with the
experimental data in Figs.~\ref{fig:1}-\ref{fig:3}. The
theoretical work also pointed out that the ac-branch should
exhibit negative slope if $U_{dc}>0$ V, for the conductive as well
as for the dielectric modes. The experiments confirmed this
behavior for the conductive EC, however, for the lowest frequency
only (inset in Fig.~\ref{fig:3}).

For higher frequencies in the conductive EC range ($f = 20$ Hz or
$f = 80$ Hz), in contrast to the theoretical expectation, the
slope of the ac-branch was experimentally found positive; i.e. in
the presence of a dc bias the EC instability sets in at higher
$U_{ac}$ values. The standard model of EC, which assumes constant
ohmic conductivity, cannot account for this finding. One should
note, however, that assuming a voltage independent (ohmic)
conductivity in the case of a weak electrolyte, such as a liquid
crystal, is not always realistic. If the applied ac voltage is not
symmetric, i. e. a nonzero dc-component is present, the number of
effective charge carriers may decrease, because a fraction of ions
is immobilized at the electrodes coated with insulating
(polyimide) surfaces. Therefore the bulk conductivity of the
liquid crystal is expected to decrease with increasing $U_{dc}$;
this was actually verified by simultaneous conductivity
measurements. Consequently, during the experiments shown in
Figs.~\ref{fig:1}-\ref{fig:3} the conductivity changes from point
to point in the $U_{ac}$-$U_{dc}$ plane, in contrast to the
constant $\sigma$ value assumed in the theoretical calculations
\cite{Krekhov2014}.

In the case of the purely ac driven conductive EC, the threshold
voltage at a fixed frequency increases if the conductivity is
reduced \cite{Bukabook1996,Peschprivate}. This increment is larger at higher
frequencies, being closer to the conductive-dielectric crossover.
This behavior offers a qualitative explanation to the shape
mismatch between the expected and measured stability limit curves
in the conductive regime. Application of $U_{dc}$ results in
lowering the conductivity, which leads to a higher onset $U_{ac}$
of EC than expected for a constant $\sigma$. The higher the
frequency, the more probable that this threshold increment flips
the slope of the ac branch from negative to positive, as found in
Figs.~\ref{fig:2} and \ref{fig:3}. Lowering the conductivity by
the dc bias reduces the crossover frequency $f_c$ as well. If due to
this reduction $f_c$ becomes lower than the driving frequency, a
dc-voltage-induced transition from the conductive to the dielectric
mode occurs, as was actually found at $f=80$ Hz (Fig.~\ref{fig:2}).

Understanding the characteristics of the dielectric EC at an
asymmetric voltage driving is more challenging. On the one hand, calculations
have shown that the ac- and dc-branches of the stability limit curve
do not connect smoothly, as there is a sharp change in the critical wave
number \cite{Krekhov2014}; these features were confirmed by the
experiments at lower frequencies in the dielectric regime (inset in
Fig.~\ref{fig:1}). On the other hand, a discrepancy exists, since in
contrast to the theory, the experimental slope of the ac-branch
is positive. The dc-voltage-induced $\sigma$ reduction (which was a clue for
the conductive EC) does not help here; in the case of dielectric EC
the theory predicts diminishing ac threshold voltages for a lower
conductivity. Consequently, when increasing $U_{dc}$, the
$U_{ac}$ component at the onset of the instability is expected to
be even smaller than without considering the change in the
conductivity, while experiments show $U_{ac}$ increasing with $U_{dc}$.
Moreover, this increment becomes larger at higher frequencies,
leading finally to the inhibition of pattern formation, i.e. the
extension of the stability limit to such $U_{ac}$, $U_{dc}$
voltage components that are several times higher than the threshold
voltages at purely ac or purely dc voltage excitation
(see the channel in Fig.~\ref{fig:1}).

The (extended) standard model of EC
is not able to account for the inhibited pattern
formation and for the dramatic effect of the frequency
on the onset characteristics of the dielectric EC at asymmetrical voltage driving.

Whatever unexpected the inhibition of electroconvective pattern formation
shown above is, it is not fully unprecedented. A
much less pronounced inhibition has already been reported
earlier \cite{Heuer2006,Pietschmann2010}
for samples driven by a
superposition of harmonic or square waves, with frequency ratios
of a small integer number, i.e. by signals
with the same time symmetries.
In that case the pattern-free region could be
extended only by a few percent of the higher $f$ ac voltage, in
contrast to the huge increment shown in Fig.~\ref{fig:1}.

We anticipate that ionic effects originating in the electrolytic
nature of liquid crystals, such as voltage dependent conductivity
(electro-purification), internal voltage
attenuation in the cell, and Debye layers at the boundaries may
play an important role. Some earlier reports indicated that the
dielectric EC rolls are rather localized at the cell surfaces than
in the bulk \cite{Gheorghiu2006,Bohatsch1999}. Taking into account
that significant electric field gradients may exist on the length
scale of the dielectric rolls in the vicinity of the electrodes,
this might account for why the behavior of dielectric EC is more
anomalous than the conductive EC.

In order to include the above mentioned effects and
therefore to give a more complete explanation of our findings, the
basic assumption of the (extended) SM on the ohmic conductivity
should be given up. A weak electrolyte model (WEM), accounting for
ionic dissociation-recombination processes, has been developed
more than a decade ago
\cite{Treiber1995,Dennin1996,Treiber1997,Treiber1998}. The WEM
introduced additional variables (the ionic concentrations) with
rate equations with two additional time scales (for the
recombination and migration times of ions), and with the relevant
(mostly unknown) material parameters (e.g. ionic mobilities). The
resulting set of partial differential equations, which is even
more complex than that of the extended SM, has only partially been
analyzed to prove the ionic origin of the Hopf bifurcation
(travelling waves) in AC-driven EC. We expect that the WEM,
generalized with inclusion of flexoelectric phenomena, would be a
proper theoretical tool to describe the dc voltage induced
phenomena, including the inhibition of the pattern formation. Such
an analysis is, however, a huge theoretical challenge for the
future.

\section*{Conclusions}

In summary, we reported convective pattern formation in a nematic
fluid induced by asymmetric voltage signals, exhibiting a rich
variety of morphological transitions. The experiments showed, in
contrast to our intuition and the predictions of the (extended)
standard model of EC, that the joint action of $ac$ and $dc$
voltages rather inhibits pattern formation than favors the
emergence of instabilities. While for the conductive EC a
qualitative explanation based on the change of conductivity could
be given, the question of why the pattern formation is largely
inhibited in the dielectric mode at high frequencies still needs
to be precisely answered in the future. The unexpected suppression
of pattern formation at high applied voltages can open new
horizons in studies of (sub)criticality or director fluctuations
in electric fields in voltage ranges where investigations were
believed to be impossible due to the occurrence of patterns or
turbulent flow of the material. Our finding also rises the
question whether analogous effects can be found in other dynamical
systems, such as isotropic EC, or shear induced turbulent
convection combined by electric fields.

\section*{ACKNOWLEDGEMENTS}

Financial support by the Hungarian Research Fund grants OTKA K81250 and
NN110672 are gratefully acknowledged. The authors are
indebted to Werner Pesch and Alexei Krekhov for fruitful
discussions.

\end{document}